\documentclass[aps,superscriptaddress,pra,showpacs]{revtex4-2}

\usepackage[utf8]{inputenc}
\usepackage[american,british]{babel}
\usepackage[T1]{fontenc}

\usepackage{graphicx,amsmath,amssymb,amsfonts,dsfont,enumitem,MnSymbol,comment,array,tabularx,physics,subcaption}
\DeclareCaptionSubType[Alph]{figure}

\makeatletter
\long\def\@makecaption#1#2{%
  \vskip\abovecaptionskip
  \sbox\@tempboxa{#1. #2}%
  \ifdim \wd\@tempboxa >\hsize
    \setlength{\parindent}{0pt}
    \setlength{\leftskip}{0pt}
    \setlength{\rightskip}{0pt}
    \setlength{\parfillskip}{0pt plus 1fil}
    #1. #2\par
  \else
    \global \@minipagefalse
    \hb@xt@\hsize{\hfil\box\@tempboxa\hfil}%
  \fi
  \vskip\belowcaptionskip}
\makeatother

\newcolumntype{P}[1]{>{\centering\arraybackslash}p{#1}}

\newcommand{\mc}[1]{\ensuremath{\mathcal{#1}}}

\usepackage{xcolor}

\newcommand{\algo}{TTM }
\newcommand{\algoE}{TTM}

\newcommand{\TT}{M}

\usepackage{geometry}

\begin{document}

\title{Tensor Train Multiplication}

\author{Alexios A.~Michailidis}
\affiliation{PlanQC GmbH, Lichtenbergstr. 8, 85748 Garching, Germany}
\author{Christian Fenton}
\affiliation{PlanQC GmbH, Lichtenbergstr. 8, 85748 Garching, Germany}
\author{Martin Kiffner}
\affiliation{PlanQC GmbH, Lichtenbergstr. 8, 85748 Garching, Germany}
\begin{abstract}
We present the Tensor Train Multiplication (TTM) algorithm for the elementwise multiplication of two tensor trains with bond dimension $\chi$.
The computational complexity and memory requirements of the \algo algorithm scale as  $\chi^3$ and  $\chi^2$, respectively. This represents
a significant improvement compared with the conventional approach, where the computational complexity scales as $\chi^4$ and  memory requirements scale as $\chi^3$.
We benchmark the \algo algorithm using flows obtained from artificial
turbulence generation and numerically demonstrate its improved runtime and memory scaling compared with the conventional approach.
The \algo algorithm paves the way towards GPU accelerated tensor network simulations of computational fluid dynamics problems with large bond dimensions
due to its dramatic improvement in memory scaling.
\end{abstract}

\maketitle
\section{Introduction\label{intro}}
Tensor network methods~\cite{schollwoeck:11} are an extremely powerful tool in theoretical quantum physics for
simulating quantum many-body systems.  This approach has enabled the simulation of physical systems that are otherwise
intractable, by removing unrealized correlations (entanglement) from the model. A simple Tensor Network (TN) example  often employed in large scale simulations are Matrix Product States or Tensor Trains (TTs). While more complicated TN
frameworks~\cite{ORUS2014117, PhysRevLett.99.220405} are under active investigation for critical or  quantum systems in two and three spatial dimensions, the all-around efficiency of TTs is hard to surpass.

Encouraged by their success, TNs  have been successfully applied to areas outside of quantum physics with the aim of achieving performance
enhancements compared to standard methods. Examples include the areas of machine learning~\cite{huggins_towards_2019,convy_mutual_2022,gao_enhancing_2022,mossi_matrix_2024,cheng_supervised_2021,stoudenmire_learning_2018,dilip_data_2022}, optimization problems~\cite{hao_quantum-inspired_2022,rams_approximate_2021,ali_traveling_2024,ali_polynomial-time_2024} and
partial differential equations for
computational fluid dynamics (CFD) applications~\cite{gourianov:22,hoelscher:24,kiffner:23,ye:22,peddinti:24,grelier:21,gourianov:24}.
CFD simulations with TNs promise to leverage simulations on large grids by efficiently compressing turbulent flows~\cite{gourianov:22}.
They can be be extended to wall-bounded flows~\cite{kiffner:23} and flows around complex objects via immersed boundaries~\cite{peddinti:24}.
TNs are also very well suited for benefiting from  specialized hardware acceleration like GPUs or TPUs. For example, it has been shown
that GPUs achieve a tenfold acceleration~\cite{hoelscher:24} of the examples studied in~\cite{gourianov:22}.
Furthermore, TNs have been used to study CFD-related problems in plasma physics~\cite{ye:22}, and the application of
TNs to high-dimensional probablility distributions has been studied in general and for CFD applications
in~\cite{grelier:21} and~\cite{gourianov:24}, respectively.

Solving computational fluid dynamics problems in TN format is different from other applications outside of
quantum physics because the underlying Navier-Stokes equations are nonlinear. More specifically, CFD simulations require
the pointwise multiplication of two velocity components for computing the advective term in the momentum equations~\cite{anderson:cfd}.
The works in~\cite{gourianov:22,kiffner:23,peddinti:24,hoelscher:24,gourianov:24} utilize the standard  pointwise multiplication algorithm for
two TT states described in~\cite{oseledets:11,Lubasch2018}. The computational complexity and memory requirement of this
algorithm scales as $\chi^4$ and $\chi^3$, respectively, where  $\chi$ is the TT bond dimension.
Other TT operations in CFD applications build on known DMRG-type algorithms for adding TT states and for
solving linear systems of equations~\cite{kiffner:23}. Since the complexity and memory requirement of these algorithms scales like $\chi^3$ and
$\chi^2$, respectively, pointwise multiplication is the bottleneck of CFD algorithms in TN format.
Moreover, the memory scaling of the conventional multiplication algorithm limits the feasability of CFD simulations
with TNs to relatively small bond dimensions and prohibits large scale simulations on GPU/TPU hardware where memory is limited.
Here we introduce the Tensor Train Multiplication (TTM)  algorithm for performing the pointwise multiplication with  $\mathcal{O}\left(\chi^3\right)$ floating point operations and
memory requirements scaling as $\chi^2$. The \algo algorithm thus scales
identically to other TT operations and paves the way towards
large-scale, tensor network based fluid dynamics simulations
on GPU devices.
We demonstrate the improved runtime and memory scaling over the conventional approach using examples from computational fluid dynamics.
This paper is organised as follows. The general TT formalism is introduced in Section~\ref{sec:methods}, where we describe the
conventional pointwise multiplication algorithm in Section~\ref{SSec:conventional}. The proposed \algo algorithm is introduced in Section~\ref{SSec:SSC}.
Numerical benchmarks using CFD data are provided in Section~\ref{sec:results}, and
a summary and discussion of our results is presented in Section~\ref{sec:summary}.
\section{Formalism and Methods}\label{sec:methods}
In this work we are interested in efficiently performing the element-wise multiplication between two TTs. In the case of vectors, this operation is just $z_j =  y_{j} x_{j}$ , which simply states that each  element of $\vec{x}$ is multiplied by the corresponding element of $\vec{y}$ and therefore requires $\text{dim}(\vec{x})$ multiplications.
Despite the simplicity of element-wise multiplication and similarly to other simple algebraic operations, an efficient TT implementation is not straightforward.

We begin by defining the TT representation of a rank-$N$ tensor $\psi \in  \mathbb{C}^{d_1\times d_2 \ldots\times d_N} $ as
\begin{equation}\label{Eq:TT}
 \psi\left(\omega^1,\ldots , \omega^N\right)=\TT^{\omega_1}\TT^{\omega_2}\cdots
\TT^{\omega_N},
\end{equation} 
where $\psi(\omega^1,\ldots , \omega^N)$ are  the complex (or real) valued tensor elements.  In this work we use two-dimensional spaces $d_i = 2$ with  $\omega_i \in \{0,1\}$ and $i \in \{1,\ldots,N\}$. Each of the $N$  carriages
$\TT^{\omega_i}\equiv \TT[i]^{\omega_i}$, graphically defined in Fig.~\ref{fig1}.A, is a matrix with dimensions $D_{i-1}\times D_{i}$, where
\begin{equation}\label{Eq:bond}
 D_i =\min\left(2^i,2^{N-i},\chi\right),
\end{equation}
and the maximal matrix dimension $\chi$ is called the bond dimension of the TT.
For clarity, we will ignore the reduction of bond dimension around the boundaries of the
TT for resource estimates in the following sections.
According to Eq.~(\ref{Eq:bond}),
the maximally allowed bond dimension is $\chi_{\text{max}}=2^{\text{floor}(N/2)}$, for which the TT representation contains
$2^N$ independent elements such that the TT represenation becomes exact. For $\chi<\chi_{\text{max}}$ the TT format
is a compressed representation of the full rank-$N$ tensor. Note that many important functions like Polynomials
and Fourier series have efficient MPS representations where the bond dimension only depends on the order of the
polynomial or the Fourier series, but not on the length $N$ of the TT~\cite{khoromskij:11,oseledets:13}. In the following we
characterize a TT by its length $N$ and bond dimension $\chi$, i.e.,
$\psi\equiv\psi[N,\chi]$.

We consider the pointwise multiplication of $\psi$ with a second TT $\phi\equiv\phi[N,\chi']$ of length $N$ and  bond  dimension $\chi'$
such that
\begin{equation}
 \Omega\left(\eta^1,\ldots , \eta^N\right) = \psi\left(\eta^1,\ldots , \eta^N\right)\phi\left(\eta^1,\ldots , \eta^N\right),
\end{equation}
where $\Omega\equiv \Omega[N,\tilde{\chi}]$ is the target TT of length $N$ and bond dimension $\tilde{\chi}$.
The TT representations of $\phi$ and $\Omega$ are
\begin{align}
 \phi\left(\sigma^1,\ldots , \sigma^N\right)& =K^{\sigma_1}K^{\sigma_2}\cdots K^{\sigma_N} ,
 \label{Eq:TTphi} \\
 \Omega\left(\eta^1,\ldots , \eta^N\right)&=\Pi^{\eta_1}\Pi^{\eta_2}\cdots \Pi^{\eta_N}, \label{Eq:TTT}
\end{align}
where the definition of the  carriages $K^{\sigma_i}$ and $\Pi^{\eta_i}$ follows the  rules  outlined below Eq.~(\ref{Eq:TT}) and
$\sigma_i,\ \eta_i \in \{0,1\}$ with $i \in \{1,\ldots,N\}$.
The task to find an efficient alogrithm that creates the TT carriages of $\Omega$
given the TT carriages of $\psi$ and $\phi$.
We first review the conventional multiplication
algorithm in Section~\ref{SSec:conventional}, and introduce the
\algo algorithm in~\ref{SSec:SSC}.
\subsection{Conventional multiplication algorithm}\label{SSec:conventional}
The conventional multiplication algorithm is best described
by noting  that we can view the
element-wise multiplication of two vectors $\vec{x}$ and $\vec{y}$
as a special case of a matrix-vector product: First, one of the vectors is elevated to a diagonal matrix, $Y_{i j} = \sum_{k}\delta^g_{i j k} y_{k}$, where the COPY tensor  $\delta^g_{i j k}$ is unity if $i = j = k$ and vanishes otherwise. To avoid confusion with the local COPY tensors defined below and in Fig.~\ref{fig1}.A we will refer to it as global COPY tensor. The matrix-vector multiplication $\vec{z} = Y \vec{x}$ results in the element-wise multiplication of the original vectors.

The conventional multiplication algorithm for TTs  $\phi$ and $\psi$  mirrors the matrix-vector approach as explained below.
  First, $\phi$ is promoted to a matrix product operator (MPO)~\cite{schollwoeck:11} $\Phi$, by applying  local COPY tensors to each physical degree of freedom,
\begin{align}
&\Phi(\sigma_1,\sigma'_1,\ldots,\sigma_N ,\sigma'_N) = K^{\sigma_1 \sigma'_1}K^{\sigma_2 \sigma'_2}\cdots K^{\sigma_N \sigma'_N},
\label{PhiMPO}\\
&K^{\sigma_n \sigma'_n}= \sum^2_{\omega_n =1}\delta_{\sigma_n \sigma'_n  \omega_n}K^{\omega_n}\notag.
\end{align} 
This operation corresponds to the application of the global COPY tensor in the uncompressed case. The application of each local COPY tensor has a cost of $\mathcal{O}\left(2^3\chi'^2\right)$, and therefore, the preparation
of the MPO in Eq.~(\ref{PhiMPO}) thus scales as $N \chi'^2$.

The contraction of $\Phi$ and $\psi$ directly results in the target TT $\Pi[N,\tilde{\chi}]$ in Eq.~(\ref{Eq:TTT})
with carriages
\begin{equation}
\Pi^{\sigma_i}_{(m m')(n n')} = \sum_{\sigma'_i}K^{\sigma_i \sigma'_i}_{m n}M^{\sigma'_i}_{m' n'}
= K_{m n}^{\sigma_i} M_{m'n'}^{\sigma_i},
\label{carriageXi}
\end{equation}
 where indices in  parentheses denote the vectorization of the multi-index tensor to a matrix. The second equality shows that $\Pi^{\sigma_i}=K^{\sigma_i}\otimes M^{\sigma_i}$ can be
 written as an outer product~\cite{oseledets:11}. The resulting TT thus has bond dimension $\tilde{\chi}= \chi  \chi'$, and the numerical complexity as well as
the memory requirements of this contraction scale as $\chi^2  \chi'^2$. However, the contraction of $\Phi$ with $\psi$
can be combined with  a variational compression~\cite{schollwoeck:11} into a TT with lower bond dimension $\tilde{\chi}\ll\chi\chi'$.
In particular, we will consider examples in Section~\ref{sec:results} where  $\text{max}(\chi,\chi')\lesssim\tilde{\chi}\ll\chi\chi'$.
The  variational compression
minimizes $\|\Omega - \Phi \psi\|^2_2$, where each tensor is sequentially optimized using a standard single-site variational algorithm~\cite{schollwoeck:11}. Assuming $\chi\sim \chi'\sim\tilde{\chi} \gg 2$, the computational cost of the variational algorithm scales as $\chi^4$.
Note that the simultaneous contraction and compression also lowers the scaling of the memory requirements from $\chi^4$ to $\chi^3$~\cite{kiffner:23}.
\subsection{\algo algorithm}\label{SSec:SSC}
\begin{figure*}[t!]
    \centering
    \includegraphics[width=0.95\columnwidth]{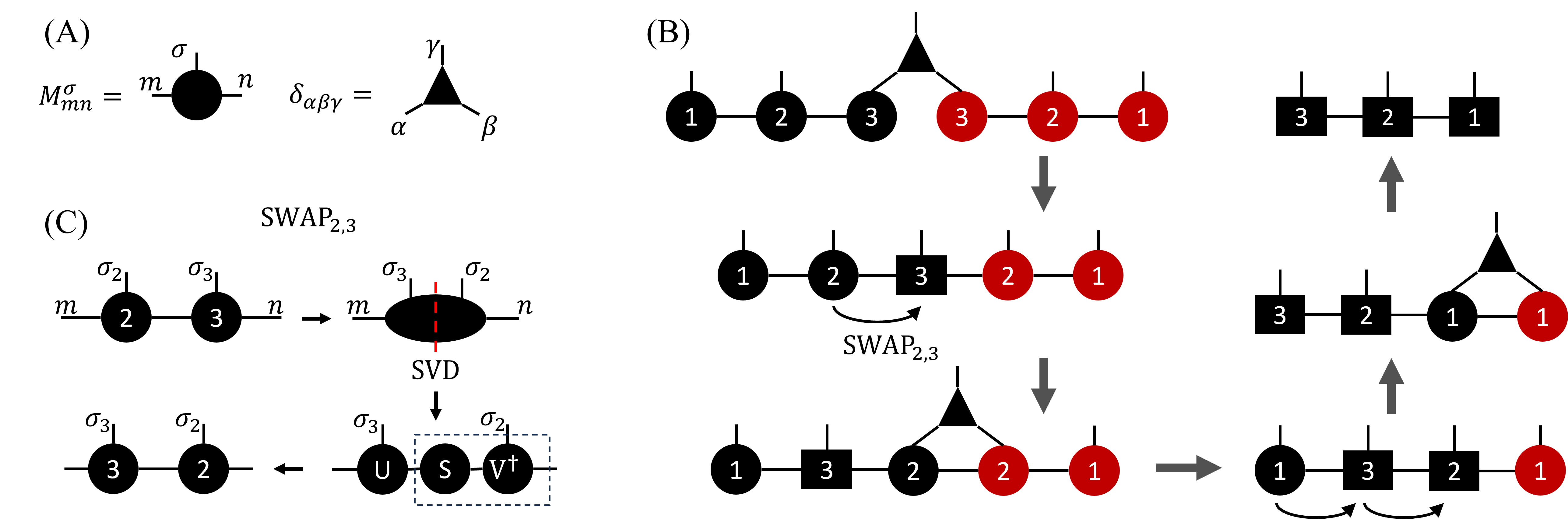}
    \caption{Illustration of the \algo algorithm. (A) Graphical definition of a tensor carriage and the local COPY tensor. (B) Pointwise multiplication  of two TTs with length $N = 3$
    using \algoE. Carriages with the same physical index are permuted to be adjacent to each other and then contracted to form the carriages of the final TT.  (C) Example of the SWAP procedure:
    Carriages 2 and 3 are  contracted and the physical indices of the resulting tensor are swapped. Next the tensor is reshaped to a  matrix by
    vectorization of indices $(m,\sigma_3)$ and $(\sigma_2,n)$.
    The new carriages are obtained by an SVD of this matrix. The singular values are absorbed to the right-most tensor to move the gauge center of the TT to the right. }
    \label{fig1}
\end{figure*}
In this section
we describe the \algo algorithm for multiplying TTs. This algorithm
is illustrated in Fig.~\ref{fig1}.B and shows a favourable scaling
with the bond dimension compared with the standard approach.
The key idea of the \algo algorithm is to avoid promoting one
TT to an MPO. Instead of implementing the multiplication as a matrix-vector operation, the strategy of \algo is to bring corresponding
 carriages of the two TTs next to each other by swapping
 operations (see Fig.~\ref{fig1}.C). This allows to form the carriage of the wanted TT by contracting the neighbouring carriages along the shared auxiliary index and their
 physical indices with the copy tensor. 
In the following, we describe each step of this algorithm  and assess the required resources.

The first step of the algorithm is to concatenate the two TTs to be
multiplied to a single TT of length $2N$,
\begin{equation}\label{Eq:tensor}
\Omega = \underbrace{K^{\sigma_1}K^{\sigma_2}\cdots 
K^{\sigma_N}}_{\phi}\underbrace{\TT^{\omega_N}\TT^{\omega_{N-1}}\cdots 
\TT^{\omega_1}}_{\psi}.
\end{equation}
This concatenation of TTs is possible because the first (last) dimension of the first (last) carriage is equal to one,
see Eq.~(\ref{Eq:bond}).
Note that we have considered the reverse order of carriages for $\psi$ in Eq.~(\ref{Eq:tensor}) in order to reduce the number of permutations as explained below. 
The carriages of
the target TT are created by iterating \textit{Contract} and \textit{Swap} operations that are described next.

\textit{Contract---} If two carriages $K^{\sigma_i}$ and $M^{\omega_i}$ with the same index $i$ are adjacent, they are contracted using a local COPY tensor to form carriage $\Pi^{\eta_{i}}$ of the target TT,
\begin{equation}\label{Eq:contract}
\Pi^{\eta_{i}}_{m n}= \sum^1_{\sigma_{i},\omega_{i}=0}\left(\sum^{\chi}_{a=1}K^{\sigma_{i}}_{m a}\TT^{\omega_{i}}_{a n}\right) \delta_{\sigma_{i}\omega_{i}\eta_{i}}.
\end{equation}
Here we have chosen the auxiliary indices to be contracted first. Alternatively, the same computational cost can be achieved by contracting one of the two carriages with the COPY tensor, and then the resulting tensor with the remaining carriage.  The computational complexity and memory requirement of the contraction is  $\mc{O}(d^2\chi^3+d^3\chi^2)$ and $\mc{O}(\chi^2)$, respectively.

\textit{Swap---} The SWAP operation is shown in Fig.~\ref{fig1}.C by example of swapping carriages 2 and 3.  After contracting the two carriages along the auxiliary index, we permute the physical legs of the resulting tensor. This operation is followed by a  singular value decomposition from
which the new carriages of the swapped representation can be obtained.
The singular value decomposition typically increases the bond dimension shared between the two carriages  by a factor of two. In order to compress the state to smaller bond dimensions,
we collect the smallest singular values in a subset
$\mc{S}\subset\Lambda$ of all singular values $\lambda_i\in\Lambda$ such that
\begin{align}
\frac{\sum_{\lambda_i \in \mathcal{S}}\lambda^2_i}{\sum_{\lambda_i\in \Lambda}\lambda^2_i}<\epsilon.
\label{Eq:eps}
\end{align}
The compressed state is obtained by only retaining the
singular values $\tilde{\Lambda} = \Lambda \setminus \mathcal{S}$.
The truncation error $\epsilon$ can be adjusted to set the precision of the swap operation, see Section~\ref{sec:results}.
In order for the compression to be globally optimal, we  bring $\Omega$ to the so called mixed canonical gauge~\cite{schollwoeck:11}. The gauge centre is set to the rightmost carriage of the pair to be swapped. In this case, the local truncation performed by Eq.~(\ref{Eq:eps}) is equivalent to a global truncation error $\|\psi- \psi'\|_2 <\sqrt{\epsilon} \|\psi\|_2$, where $\psi'$ denotes the truncated TT.

The swap and contract procedure described above is iterated until all pairs of sites are contracted.
Note that with the initial choice of $\Omega$ in
Eq.~(\ref{Eq:tensor}), the two carriages with index $N$
are already adjacent and can be contracted without further permutations.
For the $i$-th iteration the
carriage of $\psi$ must be swapped $i-1$ times in order to be adjacent to the corresponding carriage of $\phi$.
This procedure is performed by $i-1$ local SWAP operations while the TT is kept at a mixed canonical gauge centred at the rightmost site of the SWAP pair.  We note that once a contraction is performed, the gauge center needs to be moved which is achieved with QR decompositions that scale similiarly with bond dimension compared
to SVDs, but in practice, their runtime is much shorter.
\textit{Algorithmic complexity---}
In order the transform the TT in Eq.~(\ref{Eq:tensor}) to its
final form, we need to perform $N$ contractions with
numerical complexity  $\mc{O}(\chi^3)$ each. In addition, the total number of SWAP operations in the algorithm
is $N(N-1)/2$. The cost of a single swap operation requires a contraction and a  singular value decomposition which both scale as $\chi^3$.
It follows that the \algo algorithm naively scales as $\mathcal{O}\left(\chi^3 N^2\right)$. The memory requirements of the contract
and swap operations both scale like $\chi^2$, which is a
significant improvement compared to the $\chi^3$ scaling
of the conventional multiplication scheme.

The simple analysis above assumes that the bond dimension of the target TT is of the order of the bond dimension of the initial states.
As we will see in the fluid dynamics examples in Section~\ref{sec:results}, the final and initial bond dimensions are
indeed approximately the same. Of course, this holds for both conventional and \algo algorithms as it is only a property of the final state.
However, the simple analysis above ignores that the bond dimension at intermediate steps of the \algo algorithm can increase as a consequence
of swap steps. The qualitative reason for this increase is the possible presence of strong short range correlations in $\Omega$. Such correlations become long range through the swapping procedure effectively increasing the bond dimensions of the affected carriages. We emphasize that the amount of increase is problem-dependent as it depends on the properties of the TTs to be multiplied. In Appendix~\ref{sec:swapeffects} we analytically illustrate this effect using a toy model.

To summarize, we have shown that the numerical complexity of the \algo algorithm has a qubic scaling with bond dimension in contrast to the quartic scaling of the conventional algorithm. At the same time, the memory requirements of \algo scale as $\chi^2$ and therefore more favourable than
the $\chi^3$-scaling of the conventional scheme. However, the intermediate bond dimension can fluctuate depending on the multiplied TTs,
and thus the effective performance of \algo is case-dependent.
\section{Numerical benchmark results} \label{sec:results}
\begin{figure*}[t]
    \centering
     \includegraphics[width=0.85\columnwidth]{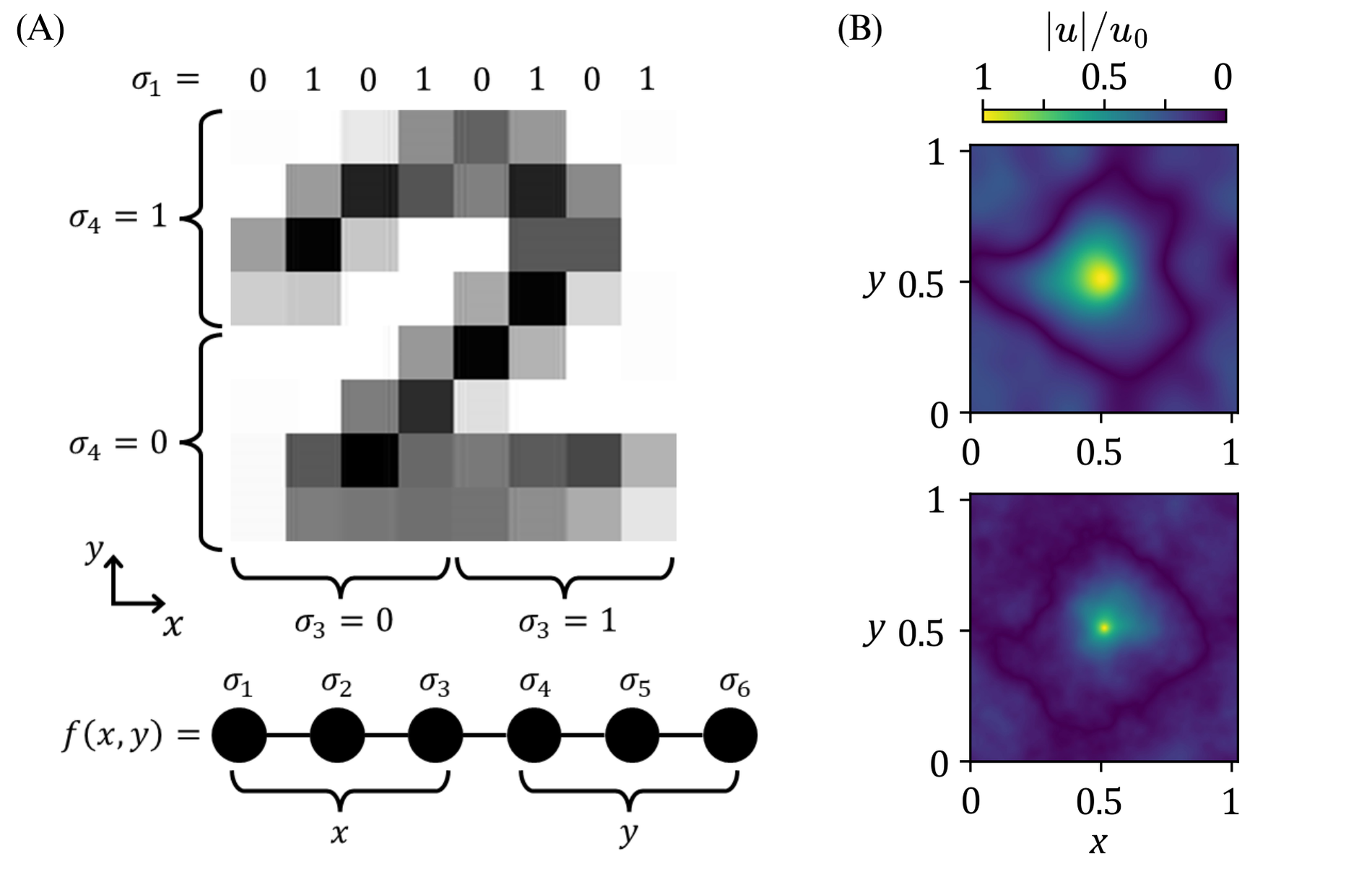}
    \caption{(A) Quantics discretization for a 2D image. Each dimension is encoded in a grid of $2^L$ vertices with $L=3$.  (B) Absolute value of velocity normalized by its maximum value  $u_0 = \mathrm{max}(\abs{u})$  of artificial turbulence fields. The bond dimension is $\chi = 41$ ($\chi = 213$) for top  (bottom) plot. 
    }
    \label{fig2}
\end{figure*}
We numerically benchmark the \algo algorithm using TT representations of flow fields from computational fluid dynamics.
More specifically, we use an artificial turbulence generation
model described in Appendix~\ref{sec:turbogen} for
the generation of turbulent flows in two spatial dimensions. We chose this example because it allows
to adjust the bond dimension over a wide range by appropriately engineering the fields in momentum space. The relation of the singular value spectrum decay to the decay of Fourier modes was recently established in the context of quantum machine learning by Jobst et al.~\cite{jobst2023efficient}. Thus, our example is ideal for studying the scaling of the algorithm with bond dimension. 
Furthermore, the performance of the \algo algorithm
for fluid flows is interesting because computational
fluid dynamics is one of its anticipated use cases. Finally,
the generated turbulent flows contain elements of randomness,
and are thus well suited as a general benchmark example.

We begin with a brief description of the so-called quantics representation~\cite{Oseledets2010,khoromskij:11} of discrete functions.
It has been recently shown that this encoding enables
a strongly compressed TT representation of flow fields even in the presence of turbulence~\cite{gourianov:22,kiffner:23}.
The quantics encoding of a real-valued function $f(x)$ with $x\in [a,b]$ is based on discretizing the interval to a uniform grid of $d^L$ points with grid spacing $\Delta x = (b-a)/(d^L-1)$. Here we assume $d = 2$. The grid points  can be encoded in different ways to the TT of Eq.~(\ref{Eq:TT}). A decomposition that has been shown to be efficient in turbulence is based on scale separation~\cite{gourianov:22}, $x = \Delta x \sum^{L}_{i=1} \sigma_i 2^{i-1} + a$, where the binary degrees of freedom $\sigma_i\in \{0,1\}$, identify the indices of the carriages in Eq.~(\ref{Eq:TT}). As it is evident, the value of $\sigma_L$ identifies the two halves of the interval, $\sigma_{L-1}$ identifies the halves of each half, etc., up to $\sigma_{1}$ which is associated to changes at the level of grid spacing $\Delta x$.

For functions of two variables $f(x,y)$, we extend the aforementioned recipe: We double the degrees of freedom $N=2 L$, while using the hierarchical encoding for both variables. In addition, we require the largest length-scales in $(x,y)$ to be adjacent, which is satisfied if we reverse the order of lengthscales in $y$,
\begin{equation}
x = \Delta x \sum^{L}_{i=1} 2^{i-1}\sigma_i  + a_x,\hspace{1 cm} y = \Delta y \sum^{L}_{i=1}2^{L-i} \sigma_{L+i}  + a_y.
\end{equation}
An illustration of the encoding is shown in Fig.~\ref{fig2}. In the following we employ the same interval for both $(x,y) \in [0,1]$, and since we use the same number of degrees of freedom for each dimension, we have $\Delta x = \Delta y = 1/(2^L-1)$.
\begin{figure*}[t!]
    \centering
     \begin{subfigure}[b]{0.45\columnwidth}
     \subcaption{ \vspace*{-1em}\hspace*{19em}}
       \centering
         \includegraphics[width=\columnwidth]{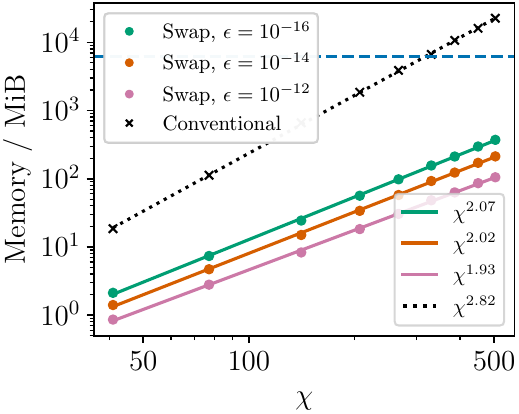}
     \end{subfigure}
     \hfill
     \begin{subfigure}[b]{0.45\columnwidth}
     \subcaption{ \vspace*{-1em}\hspace*{19em}}
       \centering
         \includegraphics[width=\columnwidth]{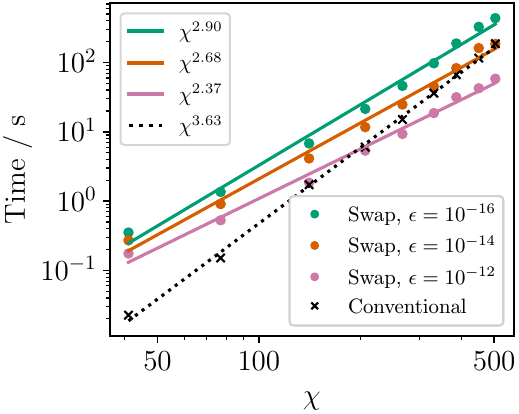}
     \end{subfigure}
     \vspace*{-1em}
     \centering
     \begin{subfigure}[t]{0.45\columnwidth}
     \centering
      \subcaption{ \vspace*{-1em}\hspace*{19em}}
         \includegraphics[width=\columnwidth]{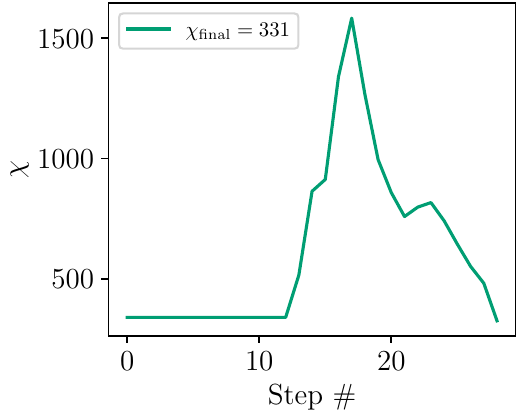}
     \end{subfigure}
     \hfill
     \begin{subfigure}[t]{0.45\columnwidth}
       \centering
      \subcaption{ \vspace*{-1em}\hspace*{19em}}
         \includegraphics[width=\columnwidth]{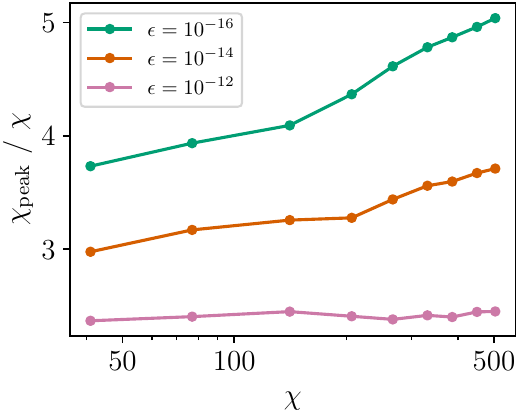}
     \end{subfigure}
     \vspace*{1em}
     \caption{Benchmarks of the \algo algorithm and performance comparison against the conventional scheme for a TT of length $N= 2L = 28$ for different precision thresholds $\epsilon$. (A) Runtime and (B) memory allocation as a function of bond dimension.(C) Maximum bond dimension of the TT following each tensor contraction for an initial bond dimension $\chi = 340$. Bond dimension significantly increases when larger length-scales are permuted at step number $\sim N$, while final bond dimension is approximately the same as the initial. (D) Increase of maximum relative bond dimension during a multiplication, as a function of the initial bond dimension of the TT.
     }
     \label{fig3}
\end{figure*}
The generation of flow fields  using
the artificial isotropic turbulence model is described in
detail  in Appendix~\ref{sec:turbogen}. Two example flow
fields are shown Figs.~\ref{fig2}.(B,C). In brief,
we generate the velocity field $\tilde{u}$  in momentum space such that it satisfies the power-law decay $\tilde{u}(k)\propto k^{-2}$, which is characteristic for turbulent flows in 2D~\cite{monin:72,monin:71}.
We employ a sigmoid function to effectively truncate the power-law by attenuating the amplitudes for wave numbers larger than a set value $k_s$. We then multiply the momentum components with random phases and transform the velocity field to real space. The
resulting function $u$ is encoded in TT format using the
quantics representation described above. We consider a  $2^L\times2^L$ grid with $L=14$ such that the TT contains $N=28$ carriages. Increasing the wave
number cutoff $k_s$ increases the TT bond dimension since turbulence features persist to finer scales [see examples in Figs.~\ref{fig2}.(B,C)].
We  generate fields with  7 different bond dimensions in the range from 50 to 500.
We then square the velocity $u$ using the  \algo
and conventional algorithms and compare the results.

First we analyse the error $\Gamma$ of the target state
$\Omega$ obtained by the TT algorithms with a reference
state $\rho$,
\begin{align}
 \Gamma = 1-\frac{\left(\Omega\cdot \rho\right)^2}{\|\Omega\|^2_2 \|\rho\|^2_2},
\end{align}
where ``$\cdot$" denotes the inner product between the two TTs.
We calculate $\rho$ by squaring every element of the unencoded
state $u$, and then encode the resulting state into a
TT~\cite{schollwoeck:11} with the same bond dimension as the target state $\Omega$.
 In case of the conventional algorithm,
$\Omega$ agrees with the reference state within machine precision.
In the case of the \algo algorithm, $\Gamma$ depends on
the truncation error $\epsilon$,
see Eq.~(\ref{Eq:eps}). We numerically find that $\Gamma$ and $\epsilon$  are  closely linked via the relation $\Gamma\approx 10 \epsilon$.
It follows that the \algo and conventional algorithms
agree within machine precision for $\epsilon\approx 10^{-16}$.

In Fig.~\ref{fig3}.A we show the  memory requirements for  the \algo algorithm and the
conventional algorithm. For this comparison, we have calculated the
exact memory requirements according to the algorithms detailed
in Section~\ref{sec:methods}.
The memory requirements of the conventional algorithm
approximately scale as $\chi^3$ as expected. Deviations are
due to boundary effects, i.e., the fact that the dimensions of
the TT carriages near the boundaries are smaller than $\chi$
[see Eq.~(\ref{Eq:bond})].
In particular, the memory requirements of
the conventional algorithm exeed the memory of the uncompressed
state for $\chi\ge 300$. In this case, the TT representation no more gives  a compression advantage.

By contrast, the memory requirements of the \algo algorithm
are up to two orders of magnitude below those of the conventional algorithm. In particular, the required memory is well below the
unencoded vector for all considered bond dimensions.
We find that the memory requirements slightly depend on
the truncation error $\epsilon$, where a larger error corresponds
to lower memory requirements. The reason for this is that
lower errors are associated with larger intermediate bond dimensions as shown below. On the other hand, the dependence of the memory
requirements on bond dimension is similar for all values of $\epsilon$ and follows closely the expected $\chi^2$ scaling. Values
of $\chi^x$ with $x>2$ are possible for low truncation errors due
to larger intermediate bond dimensions.

Figure~\ref{fig3}.B shows the runtime of the \algo and
conventional algorithms as a function of bond dimension. In the
case of the \algo alogorithm we again consider different precision thresholds $\epsilon$.  For all considered values of $\epsilon$, the scaling is significantly better than the scaling of  the conventional algorithm, with  crossover points at bond dimensions
200, 500 and 1400 for truncation errors $\epsilon$ of $10^{-12}$, $10^{-14}$ and $10^{-16}$, respectively. All numerical
experiments were carried out on a single CPU core. The scaling
of the runtime is similar to the expected result of $\chi^4$ and
$\chi^3$ for the conventional and \algo algorithm, respectively.

To better understand the dependence of the performance of
the \algo algorithm on the
truncation error, we track the largest bond dimension in between
each contraction step of the \algo algorithm in Fig.~\ref{fig3}.C.
While the bond dimensions of the velocity and the square velocity (initial and final points) are very close, the intermediate bond dimensions grow significantly. This is a numerical verification of the entanglement generation mechanism discussed in Appendix~\ref{sec:swapeffects}. To systematically understand the growth of bond dimension in the algorithm we calculate the ratio of maximum bond dimension during the algorithm to the initial bond dimension.  Figure~\ref{fig3}.D  shows that while the ratio is almost independent of the bond dimension, it depends strongly on the accuracy threshold. Since the SVD scales as $\chi^3$, the large intermediate bond dimension represents the bottleneck of the algorithm. Despite this issue,  the \algo algorithm out-scales the conventional one.


\section{Summary and discussion} \label{sec:summary}
In this work we have presented the \algo algorithm, a novel algorithm to improve the memory and runtime performance of the element-wise multiplication of TT representation of vectors. \algo is a general purpose algorithm whose performance depends only on the complexity of the functions, expressed through the singular value spectrum decay of bipartitions of the TT's. Our results illustrate a dramatic improvement in the memory cost compared to the conventional algorithm. The runtime performance depends strongly on the requested accuracy, however we have shown that for up to double precision accuracy,  \algo scales better than the conventional algorithm.

We  benchmarked the \algo algorithm with functions obtained from an artificial turbulence generator for 2D flows. Since
these functions contain elements of randomness, they serve as generic examples for testing the algorithm. The main
requirement for the algorithm to be efficient, is that the intermediate bond dimensions during the swap procedure do not grow
too much. We find through numerical experiments that this is mostly determined by how quickly the singular values of the TTs to
be multiplied decay. In this sense, the chosen example is a hard test case since the singular values of the chosen flow fields
decay only as $\lambda_i\propto i^m$, where $m \approx 2$ (see Fig.~\ref{fig4}.B in Appendix B). We also applied the \algo algorithm to other CFD data, e.g., the flows presented in~\cite{kiffner:23} and found that
the intermediate growth of bond dimension is much less pronounced.
Thus, we expect the \algo algorithm to work equally well for CFD flows in  higher dimensions
and other types of functions outside of the CFD context.

Overall, there are two directions where we believe \algo can significantly improve the simulation costs of current state-of-the-art simulations:
\begin{enumerate}
 \item[(1)] Memory-limited hardware, such as GPUs. Recently, it was shown that GPUs are significantly more efficient and scale better than CPUs at certain TT simulations~\cite{hoelscher:24,10633902,novikov2020tensor,qu2021hardware,xiang2023distributedmultigpuabinitio,menczer2024parallelimplementationdensitymatrix}. The main drawback is that the efficiency is strongly dependent on storing the relevant constituents of the simulation in the limited memory of the GPU. The \algo algorithm will allow single GPUs to perform simulations with larger bond dimensions and reduce the number of GPUs, and consequently the interconnect overhead, in multi-GPU simulations. This will enable efficient simulation of fluids at higher Reynolds numbers, which is essential for many practical applications, such as aviation and atmospheric modelling.
\item[(2)] Low precision simulations. While CFD simulations with single precision can suffer from convergence and accuracy problems, mixed-precision algorithms can often accelerate the simulations without sacrificing numerical accuracy and stability~\cite{abdelfattah2021survey}. TT implementations of such algorithms will greatly benefit from the \algo algorithm, both in runtime and memory even for low bond dimensions.

\end{enumerate}

Below we discuss how to further enhance \algo and the general TT representation of functions. The bottleneck of many TT algorithms including PDE solvers and \algo is the decay rate of the singular value spectrum of the TT. The presence of scale separation provides an efficient recipe on how to represent a single variable function, $f(x)$, by associating the order of length-scales to the order of carriages. However, for multi-parameter functions $f(x_1,x_2,\ldots)$ commonly appearing in higher dimensions, the relative positions of carriages representing different parameters requires additional optimisation. In Eq.~(\ref{Eq:tensor}) we empirically achieve this by minimizing the distance of carriages representing the largest length-scales in each dimension. However, a general purpose method for performing such a compression by permutations or more general transformations of the basis of the TT representation is not known. Research towards this direction will improve both the efficiency of practical algorithms and our understanding of the TT strength at representing functions.

\section*{Acknowledgments}
M.K. thanks Tristan Farrow for discussions. The authors acknowledge support by the European Union's Horizon Europe research and innovation program (HORIZON-CL4-2021-DIGITAL-EMERGING-02-10) under grant agreement No. 101080085 QCFD. 


\appendix

\section{Swap effects on bond dimension} \label{sec:swapeffects}
In this section we use a simple toy model to illustrate how long range permutations can affect the bond dimension of a TT. This effect is important since TTs are efficient when the correlations between bits are short range, and therefore, permutations typically lead to less compressed TTs. In this example we will employ a TT $\theta[N,\chi]$ consisting of $N>2$ carriages which has the following structure,
\begin{equation}
\TT^{\sigma_2 =0}=\left(\TT^{\sigma_1 =0}\right)^T = 2^{-1/4}(1, 0),\hspace{1 cm} \left(\TT^{\sigma_2 =1}\right)^T=\TT^{\sigma_1 =1} = 2^{-1/4}(0, 1),
\end{equation}
where $T$ denotes the transpose of the matrix. The carriages $\TT^{\sigma_n}$ with $n>2$ are arbitrary. Our definition of the first two carriages imply that the they are not correlated to the rest of the TT, i.e. $\TT^{\sigma_2}$ is a $2\times 1$ matrix. However, the first two carriages are maximally pairwise correlated as seen from the two-bit matrix,
\begin{equation}
\TT^{\sigma_1}\TT^{\sigma_2}=\begin{pmatrix} 1/\sqrt{2} & 0 \\
0 &1/\sqrt{2}
\end{pmatrix}_{\sigma_1\sigma_2},
\end{equation}
which is already in diagonal form, and therefore, its singular values are $\lambda_1  = \lambda_2 = 1/\sqrt{2}$. To explore the effects of long range swapping, we permute the second bit to the last carriage: $\sigma_1\sigma_2 \ldots  \sigma_N\rightarrow \sigma_1 \sigma_3\ldots \sigma_N \sigma_2$. We consider a bipartition of the $TT$, e.g. $\{1, j\}\cup \{j+1, N\}$ with $j>2$, and explore what happened to the bond dimension of the tensors during the swap process. For notational convenience we relabel the bits in the TT to have increasing order. We consider the matrix unfolding of the TT around the bipartition defined by the matrix elements,
 \begin{equation}\label{Eq:theta}
 \theta_{(\sigma_1 \ldots\sigma_j)(\sigma_{j+1} \ldots\sigma_N)} = \theta_{\sigma_1 \sigma_N}\otimes \theta_{(\sigma_2 \ldots\sigma_j)(\sigma_{j+1} \ldots\sigma_{N-1})},
 \end{equation}
where the tensor product structure originates from the first two carriages not being correlated with the rest of the TT. To extract the correlations between the bits defining the rows and the columns of the TT we perform the singular value decomposition of $\theta$. Using Eq.~(\ref{Eq:theta}) we find that the singular values have the form $\vec{\lambda}=(\lambda_{1},\lambda_{2})\otimes \vec{\lambda}_{\text{init}}$, where $\vec{\lambda}_{\text{init}}$ are the singular values for the same partition before the permutation. So, each singular value of the original TT becomes doubly degenerate. An intuitive comparison between the original and the final correlation structure is given by the entanglement entropy $S = -\sum_i \tilde{\lambda}_i^2 \log_2 \left(\tilde{\lambda}_i^2\right)$, where $\tilde{\lambda}_i = \lambda_i/\sqrt{\sum_i\lambda^2_i}$ are the normalized singular values. In this case we see that $S_{final} =  S_{init}+1$ revealing that an extra bit of information is required to correlate the two partitions of the TT. This information practically encodes that the value of the first bit is perfectly correlated to the value of the last bit.

While our example is very simple since the first two bits were initially not correlated with the rest, it qualitatively captures the effect of long range permutations. I.e. if the originally neighbouring bits are strongly correlated, the information of their correlation will have to pass through the intermediate TT carriages once a permutation takes place. This information will effectively increase the entanglement between different contiguous partitions in the TT, and therefore, increase the bond dimension of the intermediate carriages. Despite this drawback, in the numerical examples of the main text we observe that the \algo algorithm  scales better than the conventional one in both compute time and memory.

\begin{figure*}[t!]
    \centering
     \begin{subfigure}[b]{0.48\columnwidth}
        \subcaption{ \vspace*{-1em}\hspace*{19em}}
       \centering
         \includegraphics[width=\columnwidth]{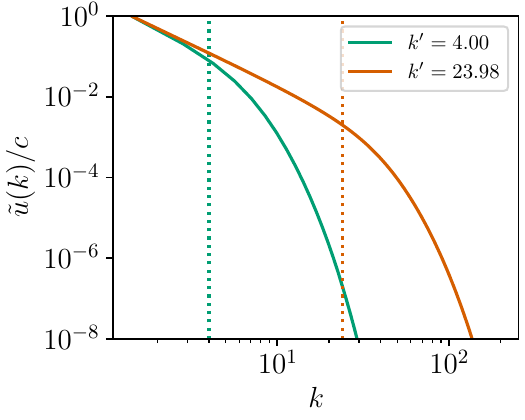}
     \end{subfigure}
     \hfill
     \begin{subfigure}[b]{0.48\columnwidth}
        \subcaption{ \vspace*{-1em}\hspace*{19em}}
       \centering
         \includegraphics[width=\columnwidth]{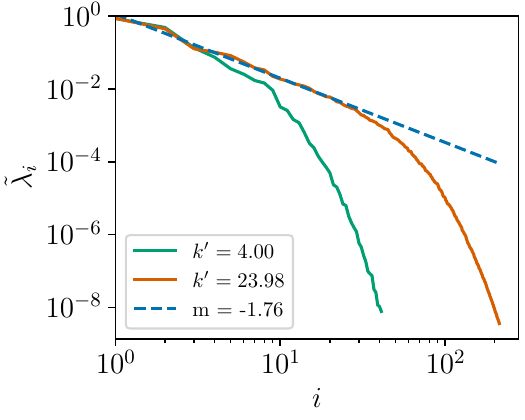}
     \end{subfigure}    
     \caption{(A) Decay of normalized velocity magnitude with momentum for the data shown in Figure~\ref{fig3} of the main text. Cutoffs $k'$ are illustrated by the dashed lines. (B) Singular values at the centre of the TTs generated by Fourier transforming the velocity profiles with magnitude decays shown in (A).  The dashed line is a power-law fit up to the cutoff momentum for the curve with the largest cutoff.  
     }
     \label{fig4}
\end{figure*}
\section{Artificial Turbulence generator} \label{sec:turbogen}
In this section we describe the algorithm used to generate the turbulent velocity fields used in the main text. We are interested in the universal part of turbulence which corresponds to the inertial range (i.e. intermediate length scales) of a turbulent flow.  The universality is expressed through the turbulent kinetic energy (TKE) spectrum, which  in two dimensions has the form
\begin{equation}\label{eq:TKE}
E(k) = \frac{1}{2}\int_0^{2\pi} |\vec{u}(k, \phi)|^2  k d\phi \propto k^{-3}, 
\end{equation}
where $k = \sqrt{k^2_x + k^2_y}$, $\phi = \tan^{-1}(k_y/k_x)$ are the polar coordinates and $\vec{u}$ is the velocity field. By disregarding the dependence of the velocity magnitude on $\phi$, Eq.~(\ref{eq:TKE}) reveals its universal decay  $|\vec{u}(k)|\propto k^{-2}$. 

Before transforming to real space, we introduce an additional function whose aim is to to control the bond dimension of the TT. This is achieved by controlling the size of the inertial range by attenuating the large momenta~\cite{jobst2023efficient} with a sigmoid function
\begin{equation}
\sigma(k) = \frac{1}{1 + \exp\left(\beta (k - k')\right)}, 
\end{equation}
where $k'$ controls the soft cut-off of the inertial range and $\beta$ controls the strength of the decay. In practice, we found $ \beta = 2 / k' $ to be a good choice for preserving the power law spectrum while attenuating amplitudes corresponding to $k > k'$.

The artificial real space velocity field is then generated straightforwardly. We define the magnitude 
\begin{equation}
\tilde{u}(k) = c k^{-2} \sigma(k).
\end{equation}
The constant $c$ is set by the normalization of the final velocity fields.
In Figure~\ref{fig4}.A we show the effects of the sigmoid function on the velocity decays $\tilde{u}(k)$. Next we use random phases in order to ensure isotropy and homogeneity.  For this, we sample random numbers $\theta_i(k_x,k_y),\in [-\pi, \pi]$ under the condition $\theta_i(k_x,k_y) = -\theta_i(-k_x,-k_y)$,
where the subscript $i\in \{x,y\}$. The swap of angle signs according to the signs of momenta guarantees that the Fourier transform of the fields is real valued due to the presence of conjugate pairs. We then define the velocity fields 
\begin{equation}
u_i (k_x,k_y) = \frac{1}{\sqrt{2}}\tilde{u}(k)e^{i\theta_i(k_x,k_y)}.
\end{equation}
Finally we perform a discrete Fourier transform and extract the velocity profiles in real space. In Figure~\ref{fig4}.B we numerically verify that the velocity profiles shown in Figure~\ref{fig4}.A correspond to TTs with singular value spectra displaying approximately the same decay profile.

In the following we summarize the details of the calculation. The flow  field is defined on the domain $\mathcal{D} \in [0,L]^2$ and discretised with $n$ evenly spaced grid points in each dimension. The allowed positions on the grid are $r_j \in [0, h, \ldots, (n - 1)h] $ for each dimension $j \in [1, 2]$, where $h = L / (n - 1)$ is the grid spacing. The minimal wavelength $\lambda$ that can be resolved on this grid without aliasing is $\lambda_{\mathrm{min}} = 2h$. For wave-vectors $k = 2 \pi / \lambda$, the maximal wave-vector is then $k_{\mathrm{max}} = \pi / h = (n - 1) \pi$.  To avoid aliasing, we set all $u_i(k > k_{\mathrm{max}}) = 0$. We also set $u_i(k < 10^{-16}) = 0$ to avoid divide-by-zero errors.

%
  

%

\begin{thebibliography}{36}%
  \makeatletter
  \providecommand \@ifxundefined [1]{%
   \@ifx{#1\undefined}
  }%
  \providecommand \@ifnum [1]{%
   \ifnum #1\expandafter \@firstoftwo
   \else \expandafter \@secondoftwo
   \fi
  }%
  \providecommand \@ifx [1]{%
   \ifx #1\expandafter \@firstoftwo
   \else \expandafter \@secondoftwo
   \fi
  }%
  \providecommand \natexlab [1]{#1}%
  \providecommand \enquote  [1]{``#1''}%
  \providecommand \bibnamefont  [1]{#1}%
  \providecommand \bibfnamefont [1]{#1}%
  \providecommand \citenamefont [1]{#1}%
  \providecommand \href@noop [0]{\@secondoftwo}%
  \providecommand \href [0]{\begingroup \@sanitize@url \@href}%
  \providecommand \@href[1]{\@@startlink{#1}\@@href}%
  \providecommand \@@href[1]{\endgroup#1\@@endlink}%
  \providecommand \@sanitize@url [0]{\catcode `\\12\catcode `\$12\catcode
    `\&12\catcode `\#12\catcode `\^12\catcode `\_12\catcode `\%12\relax}%
  \providecommand \@@startlink[1]{}%
  \providecommand \@@endlink[0]{}%
  \providecommand \url  [0]{\begingroup\@sanitize@url \@url }%
  \providecommand \@url [1]{\endgroup\@href {#1}{\urlprefix }}%
  \providecommand \urlprefix  [0]{URL }%
  \providecommand \Eprint [0]{\href }%
  \providecommand \doibase [0]{https://doi.org/}%
  \providecommand \selectlanguage [0]{\@gobble}%
  \providecommand \bibinfo  [0]{\@secondoftwo}%
  \providecommand \bibfield  [0]{\@secondoftwo}%
  \providecommand \translation [1]{[#1]}%
  \providecommand \BibitemOpen [0]{}%
  \providecommand \bibitemStop [0]{}%
  \providecommand \bibitemNoStop [0]{.\EOS\space}%
  \providecommand \EOS [0]{\spacefactor3000\relax}%
  \providecommand \BibitemShut  [1]{\csname bibitem#1\endcsname}%
  \let\auto@bib@innerbib\@empty
  \bibitem [{\citenamefont {Schollwöck}(2011)}]{schollwoeck:11}%
    \BibitemOpen
    \bibfield  {author} {\bibinfo {author} {\bibfnamefont {U.}~\bibnamefont
    {Schollwöck}},\ }\bibfield  {title} {\bibinfo {title} {The density-matrix
    renormalization group in the age of matrix product states},\ }\href
    {https://doi.org/https://doi.org/10.1016/j.aop.2010.09.012} {\bibfield
    {journal} {\bibinfo  {journal} {Annals of Physics}\ }\textbf {\bibinfo
    {volume} {326}},\ \bibinfo {pages} {96} (\bibinfo {year} {2011})},\ \bibinfo
    {note} {january 2011 Special Issue}\BibitemShut {NoStop}%
  \bibitem [{\citenamefont {Orús}(2014)}]{ORUS2014117}%
    \BibitemOpen
    \bibfield  {author} {\bibinfo {author} {\bibfnamefont {R.}~\bibnamefont
    {Orús}},\ }\bibfield  {title} {\bibinfo {title} {A practical introduction to
    tensor networks: Matrix product states and projected entangled pair states},\
    }\href {https://doi.org/https://doi.org/10.1016/j.aop.2014.06.013} {\bibfield
     {journal} {\bibinfo  {journal} {Annals of Physics}\ }\textbf {\bibinfo
    {volume} {349}},\ \bibinfo {pages} {117} (\bibinfo {year}
    {2014})}\BibitemShut {NoStop}%
  \bibitem [{\citenamefont {Vidal}(2007)}]{PhysRevLett.99.220405}%
    \BibitemOpen
    \bibfield  {author} {\bibinfo {author} {\bibfnamefont {G.}~\bibnamefont
    {Vidal}},\ }\bibfield  {title} {\bibinfo {title} {Entanglement
    renormalization},\ }\href {https://doi.org/10.1103/PhysRevLett.99.220405}
    {\bibfield  {journal} {\bibinfo  {journal} {Phys. Rev. Lett.}\ }\textbf
    {\bibinfo {volume} {99}},\ \bibinfo {pages} {220405} (\bibinfo {year}
    {2007})}\BibitemShut {NoStop}%
  \bibitem [{\citenamefont {Huggins}\ \emph {et~al.}(2019)\citenamefont
    {Huggins}, \citenamefont {Patil}, \citenamefont {Mitchell}, \citenamefont
    {Whaley},\ and\ \citenamefont {Stoudenmire}}]{huggins_towards_2019}%
    \BibitemOpen
    \bibfield  {author} {\bibinfo {author} {\bibfnamefont {W.}~\bibnamefont
    {Huggins}}, \bibinfo {author} {\bibfnamefont {P.}~\bibnamefont {Patil}},
    \bibinfo {author} {\bibfnamefont {B.}~\bibnamefont {Mitchell}}, \bibinfo
    {author} {\bibfnamefont {K.~B.}\ \bibnamefont {Whaley}},\ and\ \bibinfo
    {author} {\bibfnamefont {E.~M.}\ \bibnamefont {Stoudenmire}},\ }\bibfield
    {title} {\bibinfo {title} {Towards quantum machine learning with tensor
    networks},\ }\href {https://doi.org/10.1088/2058-9565/aaea94} {\bibfield
    {journal} {\bibinfo  {journal} {Quantum Science and Technology}\ }\textbf
    {\bibinfo {volume} {4}},\ \bibinfo {pages} {024001} (\bibinfo {year}
    {2019})}\BibitemShut {NoStop}%
  \bibitem [{\citenamefont {Convy}\ \emph {et~al.}(2022)\citenamefont {Convy},
    \citenamefont {Huggins}, \citenamefont {Liao},\ and\ \citenamefont
    {Whaley}}]{convy_mutual_2022}%
    \BibitemOpen
    \bibfield  {author} {\bibinfo {author} {\bibfnamefont {I.}~\bibnamefont
    {Convy}}, \bibinfo {author} {\bibfnamefont {W.}~\bibnamefont {Huggins}},
    \bibinfo {author} {\bibfnamefont {H.}~\bibnamefont {Liao}},\ and\ \bibinfo
    {author} {\bibfnamefont {K.~B.}\ \bibnamefont {Whaley}},\ }\bibfield  {title}
    {\bibinfo {title} {Mutual information scaling for tensor network machine
    learning},\ }\href {https://doi.org/10.1088/2632-2153/ac44a9} {\bibfield
    {journal} {\bibinfo  {journal} {Machine Learning: Science and Technology}\
    }\textbf {\bibinfo {volume} {3}},\ \bibinfo {pages} {015017} (\bibinfo {year}
    {2022})}\BibitemShut {NoStop}%
  \bibitem [{\citenamefont {Gao}\ \emph {et~al.}(2022)\citenamefont {Gao},
    \citenamefont {Anschuetz}, \citenamefont {Wang}, \citenamefont {Cirac},\ and\
    \citenamefont {Lukin}}]{gao_enhancing_2022}%
    \BibitemOpen
    \bibfield  {author} {\bibinfo {author} {\bibfnamefont {X.}~\bibnamefont
    {Gao}}, \bibinfo {author} {\bibfnamefont {E.~R.}\ \bibnamefont {Anschuetz}},
    \bibinfo {author} {\bibfnamefont {S.-T.}\ \bibnamefont {Wang}}, \bibinfo
    {author} {\bibfnamefont {J.~I.}\ \bibnamefont {Cirac}},\ and\ \bibinfo
    {author} {\bibfnamefont {M.~D.}\ \bibnamefont {Lukin}},\ }\bibfield  {title}
    {\bibinfo {title} {Enhancing generative models via quantum correlations},\
    }\href {https://doi.org/10.1103/PhysRevX.12.021037} {\bibfield  {journal}
    {\bibinfo  {journal} {Phys. Rev. X}\ }\textbf {\bibinfo {volume} {12}},\
    \bibinfo {pages} {021037} (\bibinfo {year} {2022})}\BibitemShut {NoStop}%
  \bibitem [{\citenamefont {Mossi}\ \emph {et~al.}(2024)\citenamefont {Mossi},
    \citenamefont {{\v{Z}}unkovic},\ and\ \citenamefont
    {Flouris}}]{mossi_matrix_2024}%
    \BibitemOpen
    \bibfield  {author} {\bibinfo {author} {\bibfnamefont {A.}~\bibnamefont
    {Mossi}}, \bibinfo {author} {\bibfnamefont {B.}~\bibnamefont
    {{\v{Z}}unkovic}},\ and\ \bibinfo {author} {\bibfnamefont {K.}~\bibnamefont
    {Flouris}},\ }\bibfield  {title} {\bibinfo {title} {A matrix product state
    model for simultaneous classification and generation},\ }\href@noop {}
    {\bibfield  {journal} {\bibinfo  {journal} {arXiv preprint arXiv:2406.17441}\
    } (\bibinfo {year} {2024})}\BibitemShut {NoStop}%
  \bibitem [{\citenamefont {Cheng}\ \emph {et~al.}(2021)\citenamefont {Cheng},
    \citenamefont {Wang},\ and\ \citenamefont {Zhang}}]{cheng_supervised_2021}%
    \BibitemOpen
    \bibfield  {author} {\bibinfo {author} {\bibfnamefont {S.}~\bibnamefont
    {Cheng}}, \bibinfo {author} {\bibfnamefont {L.}~\bibnamefont {Wang}},\ and\
    \bibinfo {author} {\bibfnamefont {P.}~\bibnamefont {Zhang}},\ }\bibfield
    {title} {\bibinfo {title} {Supervised learning with projected entangled pair
    states},\ }\href {https://doi.org/10.1103/PhysRevB.103.125117} {\bibfield
    {journal} {\bibinfo  {journal} {Phys. Rev. B}\ }\textbf {\bibinfo {volume}
    {103}},\ \bibinfo {pages} {125117} (\bibinfo {year} {2021})}\BibitemShut
    {NoStop}%
  \bibitem [{\citenamefont {Stoudenmire}(2018)}]{stoudenmire_learning_2018}%
    \BibitemOpen
    \bibfield  {author} {\bibinfo {author} {\bibfnamefont {E.~M.}\ \bibnamefont
    {Stoudenmire}},\ }\bibfield  {title} {\bibinfo {title} {Learning relevant
    features of data with multi-scale tensor networks},\ }\href
    {https://doi.org/10.1088/2058-9565/aaba1a} {\bibfield  {journal} {\bibinfo
    {journal} {Quantum Science and Technology}\ }\textbf {\bibinfo {volume}
    {3}},\ \bibinfo {pages} {034003} (\bibinfo {year} {2018})}\BibitemShut
    {NoStop}%
  \bibitem [{\citenamefont {Dilip}\ \emph {et~al.}(2022)\citenamefont {Dilip},
    \citenamefont {Liu}, \citenamefont {Smith},\ and\ \citenamefont
    {Pollmann}}]{dilip_data_2022}%
    \BibitemOpen
    \bibfield  {author} {\bibinfo {author} {\bibfnamefont {R.}~\bibnamefont
    {Dilip}}, \bibinfo {author} {\bibfnamefont {Y.-J.}\ \bibnamefont {Liu}},
    \bibinfo {author} {\bibfnamefont {A.}~\bibnamefont {Smith}},\ and\ \bibinfo
    {author} {\bibfnamefont {F.}~\bibnamefont {Pollmann}},\ }\bibfield  {title}
    {\bibinfo {title} {Data compression for quantum machine learning},\ }\href
    {https://doi.org/10.1103/PhysRevResearch.4.043007} {\bibfield  {journal}
    {\bibinfo  {journal} {Phys. Rev. Res.}\ }\textbf {\bibinfo {volume} {4}},\
    \bibinfo {pages} {043007} (\bibinfo {year} {2022})}\BibitemShut {NoStop}%
  \bibitem [{\citenamefont {Hao}\ \emph {et~al.}(2022)\citenamefont {Hao},
    \citenamefont {Huang}, \citenamefont {Jia},\ and\ \citenamefont
    {Peng}}]{hao_quantum-inspired_2022}%
    \BibitemOpen
    \bibfield  {author} {\bibinfo {author} {\bibfnamefont {T.}~\bibnamefont
    {Hao}}, \bibinfo {author} {\bibfnamefont {X.}~\bibnamefont {Huang}}, \bibinfo
    {author} {\bibfnamefont {C.}~\bibnamefont {Jia}},\ and\ \bibinfo {author}
    {\bibfnamefont {C.}~\bibnamefont {Peng}},\ }\bibfield  {title} {\bibinfo
    {title} {A quantum-inspired tensor network algorithm for constrained
    combinatorial optimization problems},\ }\bibfield  {journal} {\bibinfo
    {journal} {Frontiers in Physics}\ }\textbf {\bibinfo {volume} {10}},\ \href
    {https://doi.org/10.3389/fphy.2022.906590} {10.3389/fphy.2022.906590}
    (\bibinfo {year} {2022})\BibitemShut {NoStop}%
  \bibitem [{\citenamefont {Rams}\ \emph {et~al.}(2021)\citenamefont {Rams},
    \citenamefont {Mohseni}, \citenamefont {Eppens}, \citenamefont
    {Ja\l{}owiecki},\ and\ \citenamefont {Gardas}}]{rams_approximate_2021}%
    \BibitemOpen
    \bibfield  {author} {\bibinfo {author} {\bibfnamefont {M.~M.}\ \bibnamefont
    {Rams}}, \bibinfo {author} {\bibfnamefont {M.}~\bibnamefont {Mohseni}},
    \bibinfo {author} {\bibfnamefont {D.}~\bibnamefont {Eppens}}, \bibinfo
    {author} {\bibfnamefont {K.}~\bibnamefont {Ja\l{}owiecki}},\ and\ \bibinfo
    {author} {\bibfnamefont {B.}~\bibnamefont {Gardas}},\ }\bibfield  {title}
    {\bibinfo {title} {Approximate optimization, sampling, and spin-glass droplet
    discovery with tensor networks},\ }\href
    {https://doi.org/10.1103/PhysRevE.104.025308} {\bibfield  {journal} {\bibinfo
     {journal} {Phys. Rev. E}\ }\textbf {\bibinfo {volume} {104}},\ \bibinfo
    {pages} {025308} (\bibinfo {year} {2021})}\BibitemShut {NoStop}%
  \bibitem [{\citenamefont {Ali}\ \emph {et~al.}(2023{\natexlab{a}})\citenamefont
    {Ali}, \citenamefont {Delgado},\ and\ \citenamefont
    {de~Leceta}}]{ali_traveling_2024}%
    \BibitemOpen
    \bibfield  {author} {\bibinfo {author} {\bibfnamefont {A.~M.}\ \bibnamefont
    {Ali}}, \bibinfo {author} {\bibfnamefont {I.~P.}\ \bibnamefont {Delgado}},\
    and\ \bibinfo {author} {\bibfnamefont {A.~M.~F.}\ \bibnamefont {de~Leceta}},\
    }\bibfield  {title} {\bibinfo {title} {Traveling salesman problem from a
    tensor networks perspective},\ }\href@noop {} {\bibfield  {journal} {\bibinfo
     {journal} {arXiv preprint arXiv:2311.14344}\ } (\bibinfo {year}
    {2023}{\natexlab{a}})}\BibitemShut {NoStop}%
  \bibitem [{\citenamefont {Ali}\ \emph {et~al.}(2023{\natexlab{b}})\citenamefont
    {Ali}, \citenamefont {Delgado}, \citenamefont {Roura},\ and\ \citenamefont
    {de~Leceta}}]{ali_polynomial-time_2024}%
    \BibitemOpen
    \bibfield  {author} {\bibinfo {author} {\bibfnamefont {A.~M.}\ \bibnamefont
    {Ali}}, \bibinfo {author} {\bibfnamefont {I.~P.}\ \bibnamefont {Delgado}},
    \bibinfo {author} {\bibfnamefont {M.~R.}\ \bibnamefont {Roura}},\ and\
    \bibinfo {author} {\bibfnamefont {A.~M.~F.}\ \bibnamefont {de~Leceta}},\
    }\bibfield  {title} {\bibinfo {title} {Polynomial-time solver of tridiagonal
    qubo and qudo problems with tensor networks},\ }\href@noop {} {\bibfield
    {journal} {\bibinfo  {journal} {arXiv preprint arXiv:2309.10509}\ } (\bibinfo
    {year} {2023}{\natexlab{b}})}\BibitemShut {NoStop}%
  \bibitem [{\citenamefont {Gourianov}\ \emph {et~al.}(2022)\citenamefont
    {Gourianov}, \citenamefont {Lubasch}, \citenamefont {Dolgov}, \citenamefont
    {van~den Berg}, \citenamefont {Babaee}, \citenamefont {Givi}, \citenamefont
    {Kiffner},\ and\ \citenamefont {Jaksch}}]{gourianov:22}%
    \BibitemOpen
    \bibfield  {author} {\bibinfo {author} {\bibfnamefont {N.}~\bibnamefont
    {Gourianov}}, \bibinfo {author} {\bibfnamefont {M.}~\bibnamefont {Lubasch}},
    \bibinfo {author} {\bibfnamefont {S.}~\bibnamefont {Dolgov}}, \bibinfo
    {author} {\bibfnamefont {Q.~Y.}\ \bibnamefont {van~den Berg}}, \bibinfo
    {author} {\bibfnamefont {H.}~\bibnamefont {Babaee}}, \bibinfo {author}
    {\bibfnamefont {P.}~\bibnamefont {Givi}}, \bibinfo {author} {\bibfnamefont
    {M.}~\bibnamefont {Kiffner}},\ and\ \bibinfo {author} {\bibfnamefont
    {D.}~\bibnamefont {Jaksch}},\ }\bibfield  {title} {\bibinfo {title} {A
    quantum-inspired approach to exploit turbulence structures},\ }\href@noop {}
    {\bibfield  {journal} {\bibinfo  {journal} {Nature Computational Science}\
    }\textbf {\bibinfo {volume} {2}},\ \bibinfo {pages} {30} (\bibinfo {year}
    {2022})}\BibitemShut {NoStop}%
  \bibitem [{\citenamefont {H{\"o}lscher}\ \emph {et~al.}(2024)\citenamefont
    {H{\"o}lscher}, \citenamefont {Rao}, \citenamefont {M{\"u}ller},
    \citenamefont {Klepsch}, \citenamefont {Luckow}, \citenamefont
    {Stollenwerk},\ and\ \citenamefont {Wilhelm}}]{hoelscher:24}%
    \BibitemOpen
    \bibfield  {author} {\bibinfo {author} {\bibfnamefont {L.}~\bibnamefont
    {H{\"o}lscher}}, \bibinfo {author} {\bibfnamefont {P.}~\bibnamefont {Rao}},
    \bibinfo {author} {\bibfnamefont {L.}~\bibnamefont {M{\"u}ller}}, \bibinfo
    {author} {\bibfnamefont {J.}~\bibnamefont {Klepsch}}, \bibinfo {author}
    {\bibfnamefont {A.}~\bibnamefont {Luckow}}, \bibinfo {author} {\bibfnamefont
    {T.}~\bibnamefont {Stollenwerk}},\ and\ \bibinfo {author} {\bibfnamefont
    {F.~K.}\ \bibnamefont {Wilhelm}},\ }\bibfield  {title} {\bibinfo {title}
    {Quantum-inspired fluid simulation of 2d turbulence with gpu acceleration},\
    }\href@noop {} {\bibfield  {journal} {\bibinfo  {journal} {arXiv preprint
    arXiv:2406.17823}\ } (\bibinfo {year} {2024})}\BibitemShut {NoStop}%
  \bibitem [{\citenamefont {Kiffner}\ and\ \citenamefont
    {Jaksch}(2023)}]{kiffner:23}%
    \BibitemOpen
    \bibfield  {author} {\bibinfo {author} {\bibfnamefont {M.}~\bibnamefont
    {Kiffner}}\ and\ \bibinfo {author} {\bibfnamefont {D.}~\bibnamefont
    {Jaksch}},\ }\bibfield  {title} {\bibinfo {title} {Tensor network reduced
    order models for wall-bounded flows},\ }\href
    {https://doi.org/10.1103/PhysRevFluids.8.124101} {\bibfield  {journal}
    {\bibinfo  {journal} {Phys. Rev. Fluids}\ }\textbf {\bibinfo {volume} {8}},\
    \bibinfo {pages} {124101} (\bibinfo {year} {2023})}\BibitemShut {NoStop}%
  \bibitem [{\citenamefont {Ye}\ and\ \citenamefont {Loureiro}(2022)}]{ye:22}%
    \BibitemOpen
    \bibfield  {author} {\bibinfo {author} {\bibfnamefont {E.}~\bibnamefont
    {Ye}}\ and\ \bibinfo {author} {\bibfnamefont {N.~F.~G.}\ \bibnamefont
    {Loureiro}},\ }\bibfield  {title} {\bibinfo {title} {Quantum-inspired method
    for solving the vlasov-poisson equations},\ }\href
    {https://doi.org/10.1103/PhysRevE.106.035208} {\bibfield  {journal} {\bibinfo
     {journal} {Phys. Rev. E}\ }\textbf {\bibinfo {volume} {106}},\ \bibinfo
    {pages} {035208} (\bibinfo {year} {2022})}\BibitemShut {NoStop}%
  \bibitem [{\citenamefont {Peddinti}\ \emph {et~al.}(2024)\citenamefont
    {Peddinti}, \citenamefont {Pisoni}, \citenamefont {Marini}, \citenamefont
    {Lott}, \citenamefont {Argentieri}, \citenamefont {Tiunov},\ and\
    \citenamefont {Aolita}}]{peddinti:24}%
    \BibitemOpen
    \bibfield  {author} {\bibinfo {author} {\bibfnamefont {R.~D.}\ \bibnamefont
    {Peddinti}}, \bibinfo {author} {\bibfnamefont {S.}~\bibnamefont {Pisoni}},
    \bibinfo {author} {\bibfnamefont {A.}~\bibnamefont {Marini}}, \bibinfo
    {author} {\bibfnamefont {P.}~\bibnamefont {Lott}}, \bibinfo {author}
    {\bibfnamefont {H.}~\bibnamefont {Argentieri}}, \bibinfo {author}
    {\bibfnamefont {E.}~\bibnamefont {Tiunov}},\ and\ \bibinfo {author}
    {\bibfnamefont {L.}~\bibnamefont {Aolita}},\ }\bibfield  {title} {\bibinfo
    {title} {Quantum-inspired framework for computational fluid dynamics},\
    }\href@noop {} {\bibfield  {journal} {\bibinfo  {journal} {Communications
    Physics}\ }\textbf {\bibinfo {volume} {7}},\ \bibinfo {pages} {135} (\bibinfo
    {year} {2024})}\BibitemShut {NoStop}%
  \bibitem [{\citenamefont {Grelier}\ \emph {et~al.}(2019)\citenamefont
    {Grelier}, \citenamefont {Nouy},\ and\ \citenamefont {Lebrun}}]{grelier:21}%
    \BibitemOpen
    \bibfield  {author} {\bibinfo {author} {\bibfnamefont {E.}~\bibnamefont
    {Grelier}}, \bibinfo {author} {\bibfnamefont {A.}~\bibnamefont {Nouy}},\ and\
    \bibinfo {author} {\bibfnamefont {R.}~\bibnamefont {Lebrun}},\ }\bibfield
    {title} {\bibinfo {title} {Learning high-dimensional probability
    distributions using tree tensor networks},\ }\href@noop {} {\bibfield
    {journal} {\bibinfo  {journal} {arXiv preprint arXiv:1912.07913}\ } (\bibinfo
    {year} {2019})}\BibitemShut {NoStop}%
  \bibitem [{\citenamefont {Gourianov}\ \emph {et~al.}(2024)\citenamefont
    {Gourianov}, \citenamefont {Givi}, \citenamefont {Jaksch},\ and\
    \citenamefont {Pope}}]{gourianov:24}%
    \BibitemOpen
    \bibfield  {author} {\bibinfo {author} {\bibfnamefont {N.}~\bibnamefont
    {Gourianov}}, \bibinfo {author} {\bibfnamefont {P.}~\bibnamefont {Givi}},
    \bibinfo {author} {\bibfnamefont {D.}~\bibnamefont {Jaksch}},\ and\ \bibinfo
    {author} {\bibfnamefont {S.~B.}\ \bibnamefont {Pope}},\ }\bibfield  {title}
    {\bibinfo {title} {Tensor networks enable the calculation of turbulence
    probability distributions},\ }\href@noop {} {\bibfield  {journal} {\bibinfo
    {journal} {arXiv preprint arXiv:2407.09169}\ } (\bibinfo {year}
    {2024})}\BibitemShut {NoStop}%
  \bibitem [{\citenamefont {Anderson}\ and\ \citenamefont
    {Wendt}(1995)}]{anderson:cfd}%
    \BibitemOpen
    \bibfield  {author} {\bibinfo {author} {\bibfnamefont {J.~D.}\ \bibnamefont
    {Anderson}}\ and\ \bibinfo {author} {\bibfnamefont {J.}~\bibnamefont
    {Wendt}},\ }\href@noop {} {\emph {\bibinfo {title} {Computational fluid
    dynamics}}},\ Vol.\ \bibinfo {volume} {206}\ (\bibinfo  {publisher}
    {Springer},\ \bibinfo {year} {1995})\BibitemShut {NoStop}%
  \bibitem [{\citenamefont {Oseledets}(2011)}]{oseledets:11}%
    \BibitemOpen
    \bibfield  {author} {\bibinfo {author} {\bibfnamefont {I.~V.}\ \bibnamefont
    {Oseledets}},\ }\bibfield  {title} {\bibinfo {title} {Tensor-train
    decomposition},\ }\href {https://doi.org/10.1137/090752286} {\bibfield
    {journal} {\bibinfo  {journal} {SIAM Journal on Scientific Computing}\
    }\textbf {\bibinfo {volume} {33}},\ \bibinfo {pages} {2295} (\bibinfo {year}
    {2011})},\ \Eprint {https://arxiv.org/abs/https://doi.org/10.1137/090752286}
    {https://doi.org/10.1137/090752286} \BibitemShut {NoStop}%
  \bibitem [{\citenamefont {Lubasch}\ \emph {et~al.}(2018)\citenamefont
    {Lubasch}, \citenamefont {Moinier},\ and\ \citenamefont
    {Jaksch}}]{Lubasch2018}%
    \BibitemOpen
    \bibfield  {author} {\bibinfo {author} {\bibfnamefont {M.}~\bibnamefont
    {Lubasch}}, \bibinfo {author} {\bibfnamefont {P.}~\bibnamefont {Moinier}},\
    and\ \bibinfo {author} {\bibfnamefont {D.}~\bibnamefont {Jaksch}},\
    }\bibfield  {title} {\bibinfo {title} {Multigrid renormalization},\ }\href
    {https://doi.org/https://doi.org/10.1016/j.jcp.2018.06.065} {\bibfield
    {journal} {\bibinfo  {journal} {Journal of Computational Physics}\ }\textbf
    {\bibinfo {volume} {372}},\ \bibinfo {pages} {587} (\bibinfo {year}
    {2018})}\BibitemShut {NoStop}%
  \bibitem [{\citenamefont {Khoromskij}(2011)}]{khoromskij:11}%
    \BibitemOpen
    \bibfield  {author} {\bibinfo {author} {\bibfnamefont {B.~N.}\ \bibnamefont
    {Khoromskij}},\ }\bibfield  {title} {\bibinfo {title} {O (d log n)-quantics
    approximation of n-d tensors in high-dimensional numerical modeling},\
    }\href@noop {} {\bibfield  {journal} {\bibinfo  {journal} {Constructive
    Approximation}\ }\textbf {\bibinfo {volume} {34}},\ \bibinfo {pages} {257}
    (\bibinfo {year} {2011})}\BibitemShut {NoStop}%
  \bibitem [{\citenamefont {Oseledets}(2013)}]{oseledets:13}%
    \BibitemOpen
    \bibfield  {author} {\bibinfo {author} {\bibfnamefont {I.~V.}\ \bibnamefont
    {Oseledets}},\ }\bibfield  {title} {\bibinfo {title} {Constructive
    representation of functions in low-rank tensor formats},\ }\href@noop {}
    {\bibfield  {journal} {\bibinfo  {journal} {Constructive Approximation}\
    }\textbf {\bibinfo {volume} {37}},\ \bibinfo {pages} {1} (\bibinfo {year}
    {2013})}\BibitemShut {NoStop}%
  \bibitem [{\citenamefont {Jobst}\ \emph {et~al.}(2023)\citenamefont {Jobst},
    \citenamefont {Shen}, \citenamefont {Riofr{\'\i}o}, \citenamefont
    {Shishenina},\ and\ \citenamefont {Pollmann}}]{jobst2023efficient}%
    \BibitemOpen
    \bibfield  {author} {\bibinfo {author} {\bibfnamefont {B.}~\bibnamefont
    {Jobst}}, \bibinfo {author} {\bibfnamefont {K.}~\bibnamefont {Shen}},
    \bibinfo {author} {\bibfnamefont {C.~A.}\ \bibnamefont {Riofr{\'\i}o}},
    \bibinfo {author} {\bibfnamefont {E.}~\bibnamefont {Shishenina}},\ and\
    \bibinfo {author} {\bibfnamefont {F.}~\bibnamefont {Pollmann}},\ }\bibfield
    {title} {\bibinfo {title} {Efficient mps representations and quantum circuits
    from the fourier modes of classical image data},\ }\href@noop {} {\bibfield
    {journal} {\bibinfo  {journal} {arXiv preprint arXiv:2311.07666}\ } (\bibinfo
    {year} {2023})}\BibitemShut {NoStop}%
  \bibitem [{\citenamefont {Oseledets}(2010)}]{Oseledets2010}%
    \BibitemOpen
    \bibfield  {author} {\bibinfo {author} {\bibfnamefont {I.~V.}\ \bibnamefont
    {Oseledets}},\ }\bibfield  {title} {\bibinfo {title} {Approximation of
     $2^{d} \times 2^{d}$ matrices using tensor decomposition},\
    }\href@noop {} {\bibfield  {journal} {\bibinfo  {journal} {SIAM Journal on
    Matrix Analysis and Applications}\ }\textbf {\bibinfo {volume} {31}},\
    \bibinfo {pages} {2130} (\bibinfo {year} {2010})}\BibitemShut {NoStop}%
  \bibitem [{\citenamefont {Monin}\ and\ \citenamefont
    {Yaglom}(2007)}]{monin:72}%
    \BibitemOpen
    \bibfield  {author} {\bibinfo {author} {\bibfnamefont {A.~S.}\ \bibnamefont
    {Monin}}\ and\ \bibinfo {author} {\bibfnamefont {A.~M.}\ \bibnamefont
    {Yaglom}},\ }\href@noop {} {\emph {\bibinfo {title} {Statistical fluid
    mechanics: mechanics of turbulence}}},\ Vol.~\bibinfo {volume} {1}\ (\bibinfo
     {publisher} {Courier Corporation},\ \bibinfo {year} {2007})\BibitemShut
    {NoStop}%
  \bibitem [{\citenamefont {Monin}\ and\ \citenamefont
    {Yaglom}(2013)}]{monin:71}%
    \BibitemOpen
    \bibfield  {author} {\bibinfo {author} {\bibfnamefont {A.~S.}\ \bibnamefont
    {Monin}}\ and\ \bibinfo {author} {\bibfnamefont {A.~M.}\ \bibnamefont
    {Yaglom}},\ }\href@noop {} {\emph {\bibinfo {title} {Statistical fluid
    mechanics, volume II: mechanics of turbulence}}},\ Vol.~\bibinfo {volume}
    {2}\ (\bibinfo  {publisher} {Courier Corporation},\ \bibinfo {year}
    {2013})\BibitemShut {NoStop}%
  \bibitem [{\citenamefont {Liu}\ \emph {et~al.}(2024)\citenamefont {Liu},
    \citenamefont {Hong}, \citenamefont {Zhang}, \citenamefont {Tong},
    \citenamefont {Kossaifi}, \citenamefont {Wang},\ and\ \citenamefont
    {Walid}}]{10633902}%
    \BibitemOpen
    \bibfield  {author} {\bibinfo {author} {\bibfnamefont {X.}~\bibnamefont
    {Liu}}, \bibinfo {author} {\bibfnamefont {H.}~\bibnamefont {Hong}}, \bibinfo
    {author} {\bibfnamefont {Z.}~\bibnamefont {Zhang}}, \bibinfo {author}
    {\bibfnamefont {W.}~\bibnamefont {Tong}}, \bibinfo {author} {\bibfnamefont
    {J.}~\bibnamefont {Kossaifi}}, \bibinfo {author} {\bibfnamefont
    {X.}~\bibnamefont {Wang}},\ and\ \bibinfo {author} {\bibfnamefont
    {A.}~\bibnamefont {Walid}},\ }\bibfield  {title} {\bibinfo {title}
    {High-performance tensor-train primitives using gpu tensor cores},\ }\href
    {https://doi.org/10.1109/TC.2024.3441831} {\bibfield  {journal} {\bibinfo
    {journal} {IEEE Transactions on Computers}\ }\textbf {\bibinfo {volume}
    {73}},\ \bibinfo {pages} {2634} (\bibinfo {year} {2024})}\BibitemShut
    {NoStop}%
  \bibitem [{\citenamefont {Novikov}\ \emph {et~al.}(2020)\citenamefont
    {Novikov}, \citenamefont {Izmailov}, \citenamefont {Khrulkov}, \citenamefont
    {Figurnov},\ and\ \citenamefont {Oseledets}}]{novikov2020tensor}%
    \BibitemOpen
    \bibfield  {author} {\bibinfo {author} {\bibfnamefont {A.}~\bibnamefont
    {Novikov}}, \bibinfo {author} {\bibfnamefont {P.}~\bibnamefont {Izmailov}},
    \bibinfo {author} {\bibfnamefont {V.}~\bibnamefont {Khrulkov}}, \bibinfo
    {author} {\bibfnamefont {M.}~\bibnamefont {Figurnov}},\ and\ \bibinfo
    {author} {\bibfnamefont {I.}~\bibnamefont {Oseledets}},\ }\bibfield  {title}
    {\bibinfo {title} {Tensor train decomposition on tensorflow (t3f)},\
    }\href@noop {} {\bibfield  {journal} {\bibinfo  {journal} {Journal of Machine
    Learning Research}\ }\textbf {\bibinfo {volume} {21}},\ \bibinfo {pages} {1}
    (\bibinfo {year} {2020})}\BibitemShut {NoStop}%
  \bibitem [{\citenamefont {Qu}\ \emph {et~al.}(2021)\citenamefont {Qu},
    \citenamefont {Deng}, \citenamefont {Wang}, \citenamefont {Chen},
    \citenamefont {Lin}, \citenamefont {Liang}, \citenamefont {Li}, \citenamefont
    {Zhang},\ and\ \citenamefont {Xie}}]{qu2021hardware}%
    \BibitemOpen
    \bibfield  {author} {\bibinfo {author} {\bibfnamefont {Z.}~\bibnamefont
    {Qu}}, \bibinfo {author} {\bibfnamefont {L.}~\bibnamefont {Deng}}, \bibinfo
    {author} {\bibfnamefont {B.}~\bibnamefont {Wang}}, \bibinfo {author}
    {\bibfnamefont {H.}~\bibnamefont {Chen}}, \bibinfo {author} {\bibfnamefont
    {J.}~\bibnamefont {Lin}}, \bibinfo {author} {\bibfnamefont {L.}~\bibnamefont
    {Liang}}, \bibinfo {author} {\bibfnamefont {G.}~\bibnamefont {Li}}, \bibinfo
    {author} {\bibfnamefont {Z.}~\bibnamefont {Zhang}},\ and\ \bibinfo {author}
    {\bibfnamefont {Y.}~\bibnamefont {Xie}},\ }\bibfield  {title} {\bibinfo
    {title} {Hardware-enabled efficient data processing with tensor-train
    decomposition},\ }\href@noop {} {\bibfield  {journal} {\bibinfo  {journal}
    {IEEE Transactions on Computer-Aided Design of Integrated Circuits and
    Systems}\ }\textbf {\bibinfo {volume} {41}},\ \bibinfo {pages} {372}
    (\bibinfo {year} {2021})}\BibitemShut {NoStop}%
  \bibitem [{\citenamefont {Xiang}\ \emph {et~al.}(2024)\citenamefont {Xiang},
    \citenamefont {Jia}, \citenamefont {Fang},\ and\ \citenamefont
    {Li}}]{xiang2023distributedmultigpuabinitio}%
    \BibitemOpen
    \bibfield  {author} {\bibinfo {author} {\bibfnamefont {C.}~\bibnamefont
    {Xiang}}, \bibinfo {author} {\bibfnamefont {W.}~\bibnamefont {Jia}}, \bibinfo
    {author} {\bibfnamefont {W.-H.}\ \bibnamefont {Fang}},\ and\ \bibinfo
    {author} {\bibfnamefont {Z.}~\bibnamefont {Li}},\ }\bibfield  {title}
    {\bibinfo {title} {Distributed multi-gpu ab initio density matrix
    renormalization group algorithm with applications to the p-cluster of
    nitrogenase},\ }\href@noop {} {\bibfield  {journal} {\bibinfo  {journal}
    {Journal of Chemical Theory and Computation}\ }\textbf {\bibinfo {volume}
    {20}},\ \bibinfo {pages} {775} (\bibinfo {year} {2024})}\BibitemShut
    {NoStop}%
  \bibitem [{\citenamefont {Menczer}\ \emph {et~al.}(2024)\citenamefont
    {Menczer}, \citenamefont {van Damme}, \citenamefont {Rask}, \citenamefont
    {Huntington}, \citenamefont {Hammond}, \citenamefont {Xantheas},
    \citenamefont {Ganahl},\ and\ \citenamefont
    {Legeza}}]{menczer2024parallelimplementationdensitymatrix}%
    \BibitemOpen
    \bibfield  {author} {\bibinfo {author} {\bibfnamefont {A.}~\bibnamefont
    {Menczer}}, \bibinfo {author} {\bibfnamefont {M.}~\bibnamefont {van Damme}},
    \bibinfo {author} {\bibfnamefont {A.}~\bibnamefont {Rask}}, \bibinfo {author}
    {\bibfnamefont {L.}~\bibnamefont {Huntington}}, \bibinfo {author}
    {\bibfnamefont {J.}~\bibnamefont {Hammond}}, \bibinfo {author} {\bibfnamefont
    {S.~S.}\ \bibnamefont {Xantheas}}, \bibinfo {author} {\bibfnamefont
    {M.}~\bibnamefont {Ganahl}},\ and\ \bibinfo {author} {\bibfnamefont
    {O.}~\bibnamefont {Legeza}},\ }\bibfield  {title} {\bibinfo {title} {Parallel
    implementation of the density matrix renormalization group method achieving a
    quarter petaflops performance on a single dgx-h100 gpu node},\ }\href
    {https://doi.org/10.1021/acs.jctc.4c00903} {\bibfield  {journal} {\bibinfo
    {journal} {Journal of Chemical Theory and Computation}\ }\textbf {\bibinfo
    {volume} {20}},\ \bibinfo {pages} {8397} (\bibinfo {year}
    {2024})}\BibitemShut {NoStop}%
  \bibitem [{\citenamefont {Abdelfattah}\ \emph {et~al.}(2021)\citenamefont
    {Abdelfattah}, \citenamefont {Anzt}, \citenamefont {Boman}, \citenamefont
    {Carson}, \citenamefont {Cojean}, \citenamefont {Dongarra}, \citenamefont
    {Fox}, \citenamefont {Gates}, \citenamefont {Higham}, \citenamefont {Li},
    \citenamefont {Loe}, \citenamefont {Luszczek}, \citenamefont {Pranesh},
    \citenamefont {Rajamanickam}, \citenamefont {Ribizel}, \citenamefont {Smith},
    \citenamefont {Swirydowicz}, \citenamefont {Thomas}, \citenamefont {Tomov},
    \citenamefont {Tsai},\ and\ \citenamefont {Yang}}]{abdelfattah2021survey}%
    \BibitemOpen
    \bibfield  {author} {\bibinfo {author} {\bibfnamefont {A.}~\bibnamefont
    {Abdelfattah}}, \bibinfo {author} {\bibfnamefont {H.}~\bibnamefont {Anzt}},
    \bibinfo {author} {\bibfnamefont {E.~G.}\ \bibnamefont {Boman}}, \bibinfo
    {author} {\bibfnamefont {E.}~\bibnamefont {Carson}}, \bibinfo {author}
    {\bibfnamefont {T.}~\bibnamefont {Cojean}}, \bibinfo {author} {\bibfnamefont
    {J.}~\bibnamefont {Dongarra}}, \bibinfo {author} {\bibfnamefont
    {A.}~\bibnamefont {Fox}}, \bibinfo {author} {\bibfnamefont {M.}~\bibnamefont
    {Gates}}, \bibinfo {author} {\bibfnamefont {N.~J.}\ \bibnamefont {Higham}},
    \bibinfo {author} {\bibfnamefont {X.~S.}\ \bibnamefont {Li}}, \bibinfo
    {author} {\bibfnamefont {J.}~\bibnamefont {Loe}}, \bibinfo {author}
    {\bibfnamefont {P.}~\bibnamefont {Luszczek}}, \bibinfo {author}
    {\bibfnamefont {S.}~\bibnamefont {Pranesh}}, \bibinfo {author} {\bibfnamefont
    {S.}~\bibnamefont {Rajamanickam}}, \bibinfo {author} {\bibfnamefont
    {T.}~\bibnamefont {Ribizel}}, \bibinfo {author} {\bibfnamefont {B.~F.}\
    \bibnamefont {Smith}}, \bibinfo {author} {\bibfnamefont {K.}~\bibnamefont
    {Swirydowicz}}, \bibinfo {author} {\bibfnamefont {S.}~\bibnamefont {Thomas}},
    \bibinfo {author} {\bibfnamefont {S.}~\bibnamefont {Tomov}}, \bibinfo
    {author} {\bibfnamefont {Y.~M.}\ \bibnamefont {Tsai}},\ and\ \bibinfo
    {author} {\bibfnamefont {U.~M.}\ \bibnamefont {Yang}},\ }\bibfield  {title}
    {\bibinfo {title} {A survey of numerical linear algebra methods utilizing
    mixed-precision arithmetic},\ }\href
    {https://doi.org/10.1177/10943420211003313} {\bibfield  {journal} {\bibinfo
    {journal} {The International Journal of High Performance Computing
    Applications}\ }\textbf {\bibinfo {volume} {35}},\ \bibinfo {pages} {344}
    (\bibinfo {year} {2021})}\BibitemShut {NoStop}%
  \end{thebibliography}
\end{document}